\documentclass[twocolumn]{aastex631}
\usepackage[utf8]{inputenc}
\usepackage{amssymb}
\usepackage{amsmath}
\usepackage{natbib}
\usepackage{graphicx}

\begin{document}
\title{Observational Constraints on Sunyaev-Zeldovich Effect Halos Around High-z Quasars}
\author{Kyle Massingill}
\affil{School of Earth and Space Exploration, Arizona State University, Tempe, AZ 85287, USA; kyle.massingill@asu.edu}
\author{Brian Mason}
\affil{National Radio Astronomy Observatory, 520 Edgemont Road, Charlottesville, VA 22903, USA}
\author{Mark Lacy}
\affil{National Radio Astronomy Observatory, 520 Edgemont Road, Charlottesville, VA 22903, USA}
\author{Bjorn H. C. Emonts}
\affil{National Radio Astronomy Observatory, 520 Edgemont Road, Charlottesville, VA 22903, USA}
\author{Ilsang Yoon}
\affil{National Radio Astronomy Observatory, 520 Edgemont Road, Charlottesville, VA 22903, USA}
\author{Jianrui Li}
\affil{Department of Astronomy, Tsinghua University, Beijing 100084, People's Republic of China}
\author{Craig Sarazin}
\affil{Department of Astronomy, University of Virginia, P. O. Box 400325, Charlottesville, VA 22904-4325, USA}


\begin{abstract}
We present continuum observations from the Atacama Large Millimeter/submillimeter Array (ALMA) of 10 high-redshift ($2.2 \le z \le 2.7$) ultraluminous quasars (QSOs) and constrain the presence of hot, ionized, circum-galactic gas in a stacking analysis. We measure a Compton-y parameter profile with a peak value of $(1.7 \pm 1.1) \times 10^{-6}$ at a radius of $\sim50$ kpc. We compare our stacked observations to active galactic nucleus (AGN) feedback wind models and generalized Navarro-Frenk-White (gNFW) pressure profile models to constrain the wind luminosity and halo mass of the stacked QSOs. Our observations constrain the observed stack's halo mass to $<1\times 10^{13}M_{\odot}$ and the stack's feedback wind power $<1\times 10^{12}L_{\odot}$, which is $<1$\% of the bolometric luminosity of the quasar.
\end{abstract}
\section{Introduction}
\label{sec:Introduction}
The lack of super massive galaxies at low redshift (\citealt{Drory_2008}; \citealt{Treu_2005}) and the ``downsizing" \citep{Cowie_1996} of star-forming galaxies below $z\sim2$ indicates some process regulating the growth of  galaxies on long time scales. Active galactic nucleus (AGN) feedback has been postulated as the mechanism regulating galaxy formation (\citealt{Scannapieco2004}; \citealt{10.1111/j.1365-2966.2006.10519.x}) and remains the best explanation for the observed galaxy sizes. An AGN has enough energy to heat gas in the circumgalactic medium (CGM) and even possibly expel gas out of the galaxy completely (\citealt{Silk1998}; \citealt{10.1111/j.1365-2966.2006.10519.x}). It is not yet well understood what role different feedback mechanisms play in stifling galaxy growth \citep{Ostriker_2010,Fabian2012}. The relative importance of feedback mechanisms, such as jets, winds, and radiation, is still being studied.

In order to better understand AGN feedback, observations must be made to constrain the energy of the the different feedback mechanisms. Quasar winds or outflows have been detected directly in X-ray for two low-z cases (\citealt{Greene_2014}; \citealt{10.1093/mnras/stu515}). X-ray detections of outflows from high-$z$ systems where AGN are most powerful are more difficult, as the faint, diffuse X-ray emission from the winds is hard to detect against the strong point source emission from the quasar \citep[e.g.][]{2022ApJ...940...47K}. Instead, detections can be obtained 
using the Sunyaev-Zel’dovich effect (SZE) \citep{1972CoASP...4..173S} around QSOs. The thermal SZE (tSZE) is the process where cosmic microwave background (CMB) emission is distorted by traveling through hot gas, inducing an inverse Compton scattering. Bulk motion of electrons in the gas also produce a kinetic SZE (kSZE). In the case of AGN feedback, most studies focus on detecting the tSZE from the superposition of multiple wind events in the CGM of the quasar, when the tSZE is dominant \citep{Scannapieco2008}. By stacking large numbers of quasars in tSZE maps from mm-wave telescopes a number of studies have claimed to detect a significant tSZE signal from quasar winds (e.g. \citealt{Chatterjee2010}; \citealt{Ruan2015}; \citealt{Crichton2016}; \citealt{Verdier2016}; \citealt{Hall2019}; \citealt{Meinke2021}). With the large beams of the mm-wave telescopes used and uncertainties regarding the contamination of the SZ signal by dust emission from the quasar host galaxies, however, it can be hard to tell if the detections correspond simply to a combination of dust emission and the tSZE from virialized gas in the massive dark matter haloes in which the quasars reside, or whether there is an extra component of the tSZE due to quasar feedback. tSZE from hot intracluster medium has been detected in the Spiderweb radio galaxy (\citealt{Mascolo2023}).

Typical AGN exist in relatively rich environments, even at high redshift. Quasar-quasar clustering analyses \citep[e.g.][]{White2012, Timlin2018,Eftekharzadeh2019}, lensing of the Cosmic Microwave Background by quasar hosts \citep{Geach2019}, the kinematics of the warm ionized gas from quasar absorption lines \citep{Lau2018} and Ly-$\alpha$ haloes \citep{Fossati2021} all suggest that quasars at $z\sim 1-4$ lie in dark matter haloes of masses $\approx 2-10\times 10^{12}h^{-1} M_{\odot}$, consistent with galaxy groups in the local Universe. Most estimates for radio-quiet quasars are towards the lower end of that range, whereas radio-loud AGN cluster more strongly, consistent with halo masses at the upper end of that range \citep{Miley2008,Retana-Montenegro2017}. 
These relatively high mass haloes led \citet{Cen2015} and  \citet{Soergel2017} to argue that the tSZE signal from stacking studies is dominated by that from the virialized haloes of the quasars. By using a halo occupation distribution model for quasar clustering \citet{Chowdhury2017} suggested that the signal from feedback in the stacking studies made prior to 2017 is significantly higher than that expected from the virialized gas in the quasar halo only at high redshifts ($z>2.5$).

Some of the ambiguities of stacking studies can be overcome using high resolution images from mm-wave interferometers, such as ALMA. These studies enable the subtraction of emission by the quasar host and any companion or foreground/background galaxies. The morphology of the SZE signal can also be compared to, for example, that of emission line gas to determine whether the signal is associated with an outflow, or more generally with the halo in which the quasar resides. However, these observations are challenging, requiring $\stackrel{>}{_{\sim}}10$ hours of integration time on ALMA to detect a signal around even the most luminous quasars. It is also not possible to separate the tSZE from any kSZE contribution without further very deep observations at other frequencies. To date, there is only one example of an SZE detection around a single QSO, the hyperluminous (bolometric luminosity $\sim 10^{15} L_{\odot}$) quasar HE 0515-4414 \citep{Lacy2019}. 

\begin{table*} [t]
\centering
\begin{tabular}{llllll}
\hline\hline
Target     & Position                          & SNR& Redshift& log$_{10}(L_{\rm bol}/L_{\odot}$) & RMS\\ 
  & & ($\sigma$) & ($z$) &  &(Jy/beam) \\
\hline
Q0050+0051 & ICRS 00:50:21.2200, 00:51:35.000  & 14  &2.22 & 13.79&2.14e-5\\ 
\hline
Q0052+0140 & ICRS 00:52:33.6700, 01:40:40.800  & 11  &       2.30 &13.99&2.00e-5\\ 
\hline
Q0101+0201 & ICRS 01:01:16.5400, 02:01:57.400  & 13   &     2.46    &13.91    &1.91e-5\\ 
\hline
Q1227+2848 & ICRS 12:27:27.4800, 28:48:47.900  & 7    &     2.26     &13.76        &1.64e-5\\ 
\hline
Q1228+3128 & ICRS 12:28:24.9700, 31:28:37.700  & 307    &  2.22   &14.55  &2.80e-5\\ 
\hline
Q1230+3320 & ICRS 12:30:35.4700, 33:20:00.500  & 13   &    2.32   &13.66   &1.99e-5\\ 
\hline
Q1416+2649 & ICRS 14:16:17.3800, 26:49:06.200  & 21    &    2.29  &13.51 &1.86e-5\\ 
\hline
Q2121+0052 & ICRS 21:21:59.0400, 00:52:24.100  & 3 &       2.37   &13.69        &2.17e-5\\ 
\hline
Q2123$-$0050& ICRS 21:23:29.4600, $-$00:50:52.900& 7  &      2.65      &14.52          &2.21e-5\\ 
\hline
Q0107+0314 & ICRS 01:07:36.9000, 03:14:59.200  & 9     &       2.28     &13.76          &2.18e-5\\
\hline
\end{tabular}
\caption{Quasars observed in this study, their position and the signal to noise ratio (SNR) achieved in the $6^{''}$ tapered continuum ALMA 12m observations. Here SNR is defined as the peak brightness of the QSO divided by overall RMS of the image. We have estimated tapered RMS from the median of the absolute deviations from the median (MAD) of the source subtracted images. We rescaled the MAD as $RMS = k\times MAD$, where $k\approx1.4826$, so as to provide the same results as the RMS for a Gaussian distribution. $L_{bol}$ is the bolometric luminosity of the Quasar.}
\end{table*}

We therefore decided to try a different approach to constrain the tSZE signal from a sample of more typically-luminous quasars, also by using ALMA, but stacking the signal from a moderate number ($\approx 10$) of ultraluminous quasars (with bolometric luminosities $\sim 10^{14} L_{\odot}$). These quasars were selected to have extended Ly$\alpha$ nebulae around them (\citealt{Cai2019}), which allow a rough estimate of the typical halo mass in the sample to be made based on the gas dynamics (uncertain due to radiative transfer effects and possible non-gravitational motions in the Ly$\alpha$-emitting gas). Using ALMA allows us to take advantage of the capability to subtract emission from discrete sources, both in the field and associated with the quasar, whilst still building up enough signal-to-noise to make a detection or place a meaningful limit on the tSZE signal from feedback. 

In this paper we present observations of ten ultraluminous QSOs at $2.2 \le z \le 2.7$ observed by the ALMA 12m array. We constrain the Compton parameter of the observed population. We compare our measured Compton parameters to generalized Navarro–Frenk–White (gNFW) (\citealt{Arnaud2010}) profiles and spherical feedback models.

\section{Observations}
\label{sec:Observations}
Our ALMA QSO targets were observed by project 2019.1.01251.S. This was a commensal program between our continuum study of the SZE and emission-line observations of CO(4-3) and [CI] as part of the SUPERCOLD-CGM survey \citep{2023ApJ...950..180L}, which both benefited from sensitive low-surface-brightness observations with ALMA Band 4. Ten pointings were used to observe ten QSO's on the ALMA 12m array band 4 ($\sim 145\,{\rm GHz}$) with an angular resolution of \(\sim\)\(2^{''}\). Targets were observed by the 12m array for \(1-2\) hours in order to achieve an root mean square RMS of $\approx 11 \mu \rm Jy$ in the continuum. Observations of these sources were also made with the ALMA 7m array but at a much higher noise level, so we do not utilize them in our analysis here. See Table 1 for a full list of targets and the achieved signal to noise ratio of the observations. Observations were made with four $2\,\rm GHz$ spectral windows (SPWs), two containing line emission (from CO and CI) and two only containing continuum.

\subsection{Data Reduction and Calibration}
\label{sec:Data Reduction and Calibration}
The standard pipeline calibrations of the ALMA data \citep{Hunter2023} were utilized for the QSO observations. Data reduction scripts supplied with the ALMA archival data were used together with the Common Astronomy Software Application (CASA; \citealt{casa2022}) version 5.6.1-8, which includes the ALMA pipeline (\citealt{Masters2020}) version r42866. We reduced the data by first splitting off the calibrated on source observations from the full data set then flagging spectral line in each SPW. For flagging spectral lines, an image cube with a channel width of $\sim8$ MHz (base bands have a full bandwidth of $1875$ MHz) was produced for each base band. Then channels with localized emission \(4\sigma\) or more above the continuum of the channel were flagged. Flagged channels were then ignored in subsequent cleaning/imaging. All four SPWs were then combined to make a single, aggregate continuum image.

Imaging and cleaning was done using CASA task ``tclean''. We imaged an area of 60$\times$60 arcseconds centered on each quasar using a pixel scale of $0.4^{''}$/pixel. We started the imaging process by first cleaning the ALMA 12m images at full resolution. We cleaned using a robustness of $0.5$. For each of the ten targets, the sources (the QSO and any nearby serendipitous sources) were masked manually; we then cleaned the masked region to the threshold of the overall image RMS. All subsequent analysis was performed upon the clean residual images, which have effectively had the central QSO (and any strong, positive emission) subtracted. We applied Gaussian tapers to the (u,v)-data during imaging to improve surface brightness measurements of extended signals. We made tapered images for each source at $\sim 6^{''}$ and $\sim 3^{''}$ resolutions.

The object Q1228+3128 (previously identified as a radio loud QSO in \citealt{Li2021}) was detected with very high (\(\sim 307\sigma\)) signal-to-noise ratio (SNR) in the ALMA continuum observations such that the image was dynamic range limited and benefited from self-calibration. To self-calibrate we use the central source to correct for changes in phase. We solved for phase in intervals of scan length following ALMA NAASC recommended self-calibration procedure (\citealt{Brogan2018}). However, even with self-calibration we determined the observation to not have enough dynamic range to detect the $\mu$Jy level signal of the tSZE. Q1228+3128 was therefore excluded from the analyses. Our second highest QSO SNR detection in the continuum observations was Q1416+2649 with a much lower significance of \(\sim 21\sigma\) (we do not utilize self-calibration for this Q1416+264). Continuum emission from the central QSO was clearly detected in 9 of the 10 targets, and weakly ($\sim 3 \sigma$) detected in the other.

\begin{figure}
    \centering
    \includegraphics[width=1\linewidth]{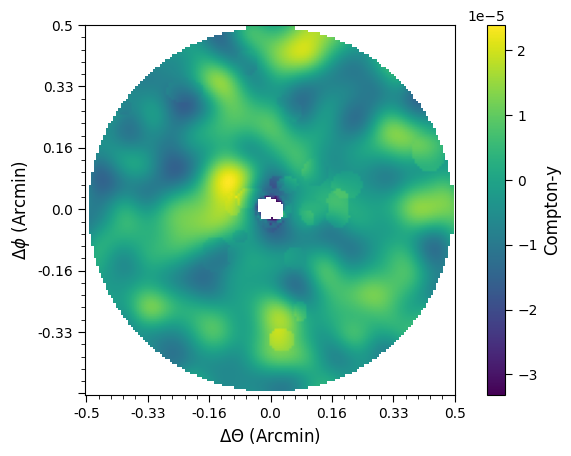}
    \caption{Stacked QSO environments expressed as a Compton-y map. A $6^{''}$ Gaussian taper has been applied to the ALMA 12m observations. Regions with source flux have been masked. Bins of differing radius have been made, with the width of each bin being about $\sim 6.7^{''}$ wide.}
    \label{fig:stacked_y}
\end{figure}

\subsection{Stacked SZE Signal}
\label{sec:SZE Signal}
We stacked the source subtracted ALMA images by first measuring the median absolute deviation (MAD) of each residual image and applying weighting to each residual image based on the MAD. The masks from the cleaning stage (see section {\ref{sec:Data Reduction and Calibration} Data Reduction and Calibration}) were applied to the residuals and masked regions were not included in the stack. This was done to ensure any leftover source emission was not added to the final stacked image. All the residuals were then stacked by summation in units of Jy/beam. We produced a $\sim 6^{''}$ and $\sim 3^{''}$ resolution stack.

We interpret the stacked continuum flux observations as SZE signal by converting to Compton parameter ($y$). Observed decrement in the continuum flux will correspond to a positive Compton parameter. Our stacking analysis will not be sensitive to the non-spherically symmetric kSZE. For stacked QSO environments we use a pure thermal SZE model to determine the Compton parameters. We therefore assume \(v_r/c<<(k_BT_e)/(m_ec^2)f_1(x)\). Following \citealt{Sazonov1998}, the change in intensity of the CMB due to SZE is given by:

\begin{equation}
    \Delta I(x) = \frac{2k_B T_{CMB}}{\lambda^2}\frac{x^2e^x}{(e^x-1)^2}\tau\frac{ k_BT_e}{m_ec^2}f_1(x),
\end{equation}
where \(x\equiv h\nu/k_BT_{CMB}\), and \(f_1(x)=x\coth(x/2)-4\). \(\tau\) is the Thompson scattering optical depth along the line of sight and \(f_1(x)\) is the frequency dependence of thermal SZE. We therefore describe the spectral distortion along a line of sight as the Compton parameter (\citealt{1972CoASP...4..173S}):
\begin{equation}
    y=\frac{\tau k_BT_e}{m_ec^2}.
\end{equation}
We convert observed flux into Compton parameter for the purposes of comparing observations to theoretical models.

Fig. \ref{fig:stacked_y} shows the Compton parameter of the stacked QSO environments tapered to an effective beam of $\sim 6''$. A region of high $y$ can be seen in the east side of the image, at a radius of $\sim 5''$ from the center, peaking at   $y= 2.4e-5$ or $\sim2.6\sigma$. Regions of null pixels exist (prominently seen as white pixels at the center of Fig. \ref{fig:stacked_y}) due to masking as described earlier in this section.

\section{Analysis}
\subsection{gNFW Models}
\label{sec:GNFW Models}
In order to compare our observations to theoretical models of different halo masses, we use a generalized Navarro-Frenk-White (gNFW) pressure profile:
\begin{equation}
\mathbb{P} (x) = \frac{P_0}{(c_{500}x)^\gamma[1+(c_{500}x)^\alpha]^{(\beta-\gamma)/\alpha}}
\end{equation}
as described in \citealt{Arnaud2010} and \citealt{Nagai2007}. $x\equiv r/r_s$ and $ r_s=r_{500}/c_{500}$, where $r_{500}$ is the radius containing matter at 500 times the ambient density. We treat the halo as a cube divided into cells; the cube is \(300^3\) cells. Each is cell is \(0.4^{''}\) per side and the total volume of the cube, within which we are integrating the line-of-sight pressure, is \(\sim 1\times 10^{10}\) kpc$^3$at our average redshift of \(z=2.34\). We then calculate the pressure of each cell:

\begin{equation}
P(r)=P_{500} \left[\frac{M_{500}}{3\times10^{14}h_{70}^{-1}M_\odot} \right]^{\alpha_P+\alpha'_P(x)}\mathbb{P} (x)
\end{equation}with parameters:
\begin{eqnarray}
&& [P_0,c_{500},\gamma,\alpha,\beta] = \nonumber \\
&& \left[ 8.403 h^{-3/2}_{70}, 1.177, 0.3081, 1.0510, 5.4905 \right]
\end{eqnarray}

where \(P_{500}\) and \(M_{500}\) are the corresponding pressure and mass respectively. To determine \(y\) we integrate the pressure through the cube. We generate halo simulation images (see section \ref{sec:Simulating Observations}) for halo masses of \(1 \times 10^{12}M_{\odot}\) , \(3 \times 10^{12}M_{\odot}\), \(1 \times 10^{13}M_{\odot}\), and \(3 \times 10^{13}M_{\odot}\). For the nominal mass case of  \(3 \times 10^{12}M_{\odot}\), $y\sim 10^{-6}$ out to a radius of $\sim 100$~kpc (or $\approx 12^{''}$ at \(z=2.34\)) from the center of the profile.

\subsection{Feedback Models}
\label{sec:Feedback Models}
The gNFW models do not include the effects of feedback. We therefore constructed a set of simple spherical wind models of AGN feedback using the prescription of \citet{2011MNRAS.412..905R} as adapted for AGN winds by \citet{Lacy2019}. These give the radius of the bubble, $R$:

\begin{equation}
    R=\beta \left(\frac{L_W T^3}{\Omega_{b}\rho_{c}(1+z)^3(1+b_{Q}\delta)}\right)^{1/5}
\end{equation}
where $\beta=0.8828$, $L_W$ is the wind power, $T$ is the age of the outflow, $\Omega_b$ is the fraction of the critical density of the Universe $\rho_c$ in baryons, $b_Q\approx 13$ is the cosmological bias factor for quasars and $\delta=180$ is the assumed overdensity corresponding to collapsed structures. The pressure inside the bubble is approximately constant and is:
\begin{equation}
    P_{\rm bub}=\frac{2}{5}\frac{3}{4 \pi}\frac{L_W T}{R^3}
\end{equation}
and the peak Compton-$y$ signal (through the center of the bubble):
\begin{equation}
    y_{\rm max}=1.08\frac{P_e\sigma_T}{m_ec^2}2R
\end{equation}
where the electron pressure, $P_e=P_{\rm bub}/1.92$, $\sigma_T$ is the Thompson cross-section and $m_e$ the electron mass. We constructed 2-D models of the SZE signal to then run through the simulator, as described below.

\begin{figure}
    \centering
    \includegraphics[width=1.0\linewidth]{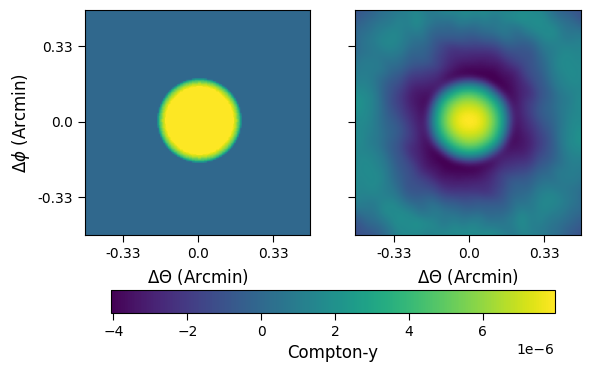}
    \caption{This is an example of a feedback model ($T = 1 \times 10^{8}$years and $L_W = 1 \times 10^{12} L_{\odot}$). LEFT: The wind model calculated as described in section \ref{sec:Feedback Models}. RIGHT: The wind model convolved with the ALMA beam using the simulator described in section \ref{sec:Simulating Observations}. In the right hand image there are primary beam and sidelobe effects such as the negative Compton-y ring at a radius of $\sim 0.16^{'}$.}
    \label{fig:simulation_demo}
\end{figure}

\begin{figure}
    \centering
    \includegraphics[scale=0.57]{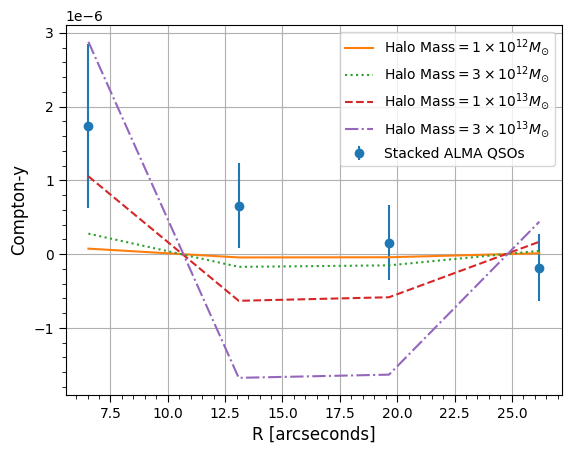}
    \caption{Radial profile analysis comparing the stacked QSO images to gNFW profiles of four different halo masses. Both the real and theoretical observations contain primary beam effects from the telescope configuration. Uncertainty in the bin mean is $\sigma = RMS/\sqrt{beams - 1}$, where ``beams'' is the number of ALMA beams in the bin.}
    \label{fig:Halo Mass RP}
\end{figure}

\subsection{Simulating Observations}
\label{sec:Simulating Observations}
The gNFW pressure profiles and wind models are simulations of sky images. In order to compare them to our observations in the image plane, we had to add the effects of the ALMA beam to our simulated images. We do this by utilizing the ALMA simulator, a facility within CASA (\citealt{casa2022}). The ALMA simulator takes a sky image and a given set of observational conditions and produces the expected ALMA visibilities. This is done by convolving the telescope configuration with the theoretical sky image. We used simulator task ``simpredict" from CASA version 5.6.1-8. For each of calculated models, we produce a set of theoretical visibilities based on the QSO observational conditions. These simulated visibities are then imaged using the same parameters as the QSOs (see Fig. \ref{fig:simulation_demo} for an example). These imaged models now have the same primary beam and side lobe effects as the observations.

\subsection{Radial Profile Analysis}
\label{Radial Profile Analysis}

To compare the gNFW models to our observations, the Compton parameter of the stack was analyzed as a radial profile in bins slightly wider then the width of the tapered beam. We utilized the lower resolution stack for this analysis, as the pressure profiles are relatively extended ( $>10^{''}$) sources. The beam of the smoothed ALMA image is $\sim 6^{''}$ while we used $\sim 6.7^{''}$ wide bins. The central pixels of the image (out to $r\sim 3^{''}$) were not utilized in the radial profile analysis as that region is mostly masked out. We measured the mean Compton parameter and RMS in each bin of the stack. Figure \ref{fig:Halo Mass RP} shows our radial profile analysis comparing the stacked QSO images to gNFW profiles of different halo masses.

\begin{figure}
    \centering
    \includegraphics[width=1\linewidth]{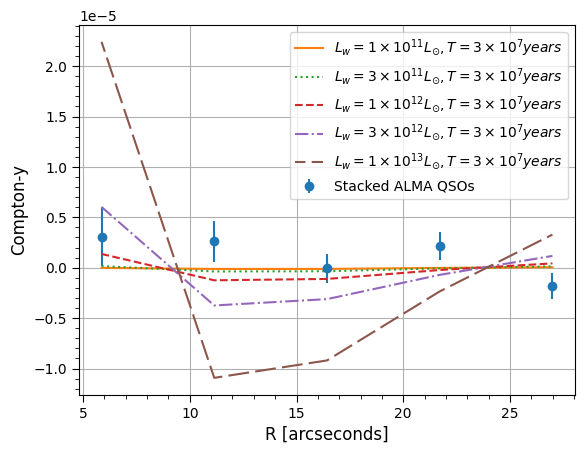}
    \caption{Radial profile analysis comparing the stacked QSO images to feedback models with an outflow age of $T = 3\times 10^{7}$ years. Both the real and theoretical observations contain primary beam effects from the telescope configuration. Uncertainty in the bin mean is $\sigma = RMS/\sqrt{\text{beams} - 1}$, where ``beams'' is the number of ALMA synthesized beams in the bin.}
    \label{fig:T=3e7 Wind RP}
\end{figure}

\begin{figure}
    \centering
    \includegraphics[width=1\linewidth]{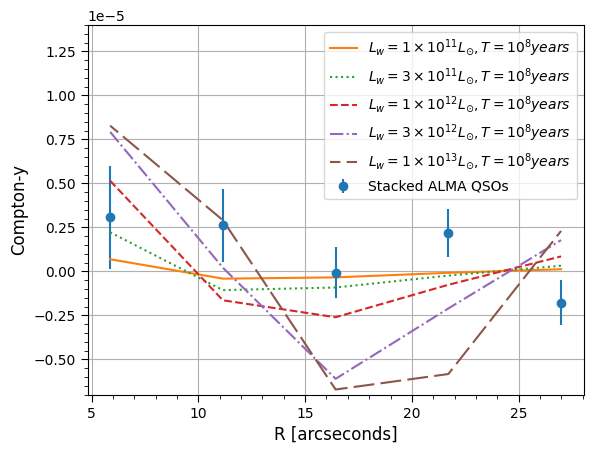}
    \caption{Radial profile analysis comparing the stacked QSO images to feedback models with an outflow age of $T = 1\times 10^{8}$ years. Both the real and theoretical observations contain primary beam effects from the telescope configuration. Uncertainty in the bin mean is $\sigma = RMS/\sqrt{\text{beams} - 1}$, where ``beams'' is the number of ALMA synthesized beams in the bin.}
    \label{fig:T=1e8 Wind RP}
\end{figure}

The feedback models were prepared using the same procedure as the gNFW models (see section \ref{sec:Simulating Observations}). We compared the feedback models to the higher resolution stack ($\sim 3^{''}$ beam) in order to be sensitive to wind bubbles with radii on order of a few arcseconds predicted by recent simulations \citep{Chakraborty2023}. We analyzed this stack in Compton-$y$ as a radial profile using bins with a width of $\sim 5.3^{''}$. As on the lower resolution image, the central pixels of the image (out to $r\sim 3^{''}$) were not utilized in the radial profile analysis. We compared this profile to the feedback models at a variety of wind powers ($L_W$) and two outflow ages ($T$). The chosen outflow ages are based on the estimated age of a previously detected feedback wind bubble in \citealt{Lacy2019} and are of the order of the Salpeter timescale for quasar growth \citep[e.g.][]{2013BASI...41...61S}. Figures \ref{fig:T=3e7 Wind RP} and \ref{fig:T=1e8 Wind RP} show our radial profile analysis.

\section{Results and Discussion}
\label{sec:Results and Discussion}

\begin{table}[]\label{tab:table_halo_mass}
\begin{tabular}{lll}
\hline\hline
Model Halo Mass              & $\chi^2$ & p-value  \\ \hline
$1 \times 10^{12} M_{\odot}$ & $4.03$       & $0.4$      \\ \hline
$3\times 10^{12} M_{\odot}$  & $4.38$        & $0.36$     \\ \hline
$1\times 10^{13}M_{\odot}$   & $8.02$        & $0.09$     \\ \hline
$3\times 10^{13}M_{\odot}$   & $31.43$       & $2.51\times 10^{-6}$ \\ \hline
\end{tabular}
\caption{Results of image plane analysis, comparing stacked observations of QSO environments to gNFW halo profiles at four different halo masses as a goodness of fit $\chi^2$ and associated statistical p-value with four degrees of freedom. All models use a concentration of 1.}
\end{table}

\subsection{Halo Mass}
\label{sec: Halo Mass}
Signal from a gNFW halo (CMB decrement) is strongest at the center of the profile, which would not be detectable by these observations as they are centered on continuum bright QSOs. On the image plane we instead try to fit the regions around the QSO to the profile the same distance from the center. In our stacked data we do see a Compoton signal in the inner radial bin (see figure \ref{fig:Halo Mass RP}), but the error is large due to the small radius; this bin only contains $\sim 10$ beams. The other notable features of the gNFW halo when observed through the ALMA simulator are the dark sidelobes. Larger mass halo models will have brighter (in Compton-$y$) centered decrement and darker sidelobes than lower mass models. Our observations lack dark sidelobes that would indicate a higher mass halo. We quantify the relationship between the stacked QSO observations and the simulated gNFW models by calculating the $\chi^2$ and p-value (where ``p'' is the probability that the difference between model and observations are due to chance) for each halo mass model compared to the stacked observations. $\chi^2$ is calculated from 4 bins in the case of halo mass. Looking at Table 2 we can see that as we go to higher halo mass models, it becomes statistically less likely that they are consistent with our observations. The gNFW model with halo mass $1\times 10^{13}M_{\odot}$ has a p-value $0.09$. We can therefore constrain our observed stack of QSOs to have a halo mass $<1\times 10^{13}M_{\odot}$ with $90\%$ confidence.

\begin{table}[]\label{tab:wind1}
\begin{tabular}{lll}
\hline\hline
\multicolumn{3}{c}{$T = 3 \times 10^{7}$years}                                                                \\ \hline
\multicolumn{1}{l}{$L_{W}$}                    & \multicolumn{1}{l}{$\chi^2$} & p-value              \\ \hline
\multicolumn{1}{l}{$1 \times 10^{11} L_{\odot}$} & \multicolumn{1}{l}{$1.49$}      & $0.19$               \\ \hline
\multicolumn{1}{l}{$3 \times 10^{11} L_{\odot}$} & \multicolumn{1}{l}{$1.6$}       & $0.16$               \\ \hline
\multicolumn{1}{l}{$1 \times 10^{12} L_{\odot}$} & \multicolumn{1}{l}{$2.08$}      & $0.06$               \\ \hline
\multicolumn{1}{l}{$3 \times 10^{12} L_{\odot}$} & \multicolumn{1}{l}{$4.93$}      & $1.6 \times 10^{-4}$ \\ \hline
\multicolumn{1}{l}{$1 \times 10^{13} L_{\odot}$} & \multicolumn{1}{l}{$30.56$}     & $<1 \times 10^{-16}$ \\ \hline
\end{tabular}
\caption{Results of image plane analysis, comparing stacked observations of QSO environments to outflow age $3 \times 10^{7}$~years feedback models at five different wind powers as a goodness of fit $\chi^2$ and associated statistical p-value with five degrees of freedom.}
\end{table}

\begin{table}[]\label{tab:wind2}
\begin{tabular}{lll}
\hline\hline
\multicolumn{3}{c}{$T = 1 \times 10^{8}$years}                                                                  \\ \hline
\multicolumn{1}{l}{$L_{W}$}                    & \multicolumn{1}{l}{$\chi^2$} & p-value                \\ \hline
\multicolumn{1}{l}{$1 \times 10^{11} L_{\odot}$} & \multicolumn{1}{l}{$1.55$}      & $0.17$                 \\ \hline
\multicolumn{1}{l}{$3 \times 10^{11} L_{\odot}$} & \multicolumn{1}{l}{$1.88$}      & $0.09$                 \\ \hline
\multicolumn{1}{l}{$1 \times 10^{12} L_{\odot}$} & \multicolumn{1}{l}{$3.32$}      & $0.01$                 \\ \hline
\multicolumn{1}{l}{$3 \times 10^{12} L_{\odot}$} & \multicolumn{1}{l}{$7.78$}      & $2.51 \times 10^{-7}$  \\ \hline
\multicolumn{1}{l}{$1 \times 10^{13} L_{\odot}$} & \multicolumn{1}{l}{$13.68$}     & $2.18 \times 10^{-13}$ \\ \hline
\end{tabular}
\caption{Results of image plane analysis, comparing stacked observations of QSO environments to outflow age $1 \times 10^{8} $ years feedback models at five different wind powers as a goodness of fit $\chi^2$ and associated statistical p-value with five degrees of freedom.}
\end{table}

\subsection{Feedback}
\label{sec:Feedback}
We have modeled feedback as a spherical wind bubble that will cause a positive Compton-$y$ signal within the radius of the bubble. Sidelobe effects from the ALMA observatory are accounted for by the simulator and cause negative Compton-$y$ signal around the feedback wind bubble. On the image plane we fit the regions around the QSO to the feedback models at the same distance from image center. We quantify the relationship between the stacked QSO observations and the simulated feedback models by calculating the $\chi^2$ and p-value for each wind power and outflow age model compared to the stacked observations. Looking at Tables 3 and 4 we see that in both outflow age cases, as we go to higher power models, it becomes statistically less likely that they are consistent with our observations. At a wind power of $1 \times 10^{12} L_{\odot}$ and an outflow age $1 \times 10^{8}$~years  the p-value is $0.01$ We can therefore constrain our observed stack of QSOs to have a feedback wind power $<1\times 10^{12}L_{\odot}$ , or $\stackrel{<}{_{\sim}}1$\% of the bolometric luminosity of the quasar.

\section{Conclusions}
\label{sec:Conclusions}
We use the constraints on the tSZE from our stack of quasar observations to show that (1) if the quasars are in quiescent halos of virialized gas, the halo mass is $<1\times 10^{13}M_{\odot}$ (consistent with estimates $\sim 3\times 10^{12}M_{\odot}$ from clustering analyses), and (2), if feedback from thermal winds extends to spatial scales $\sim 100$~kpc, these winds carry $<1\%$ of the bolometric luminosity of the quasars. This finding is consistent with the only direct detection of SZE around a single QSO, \citealt{Lacy2019}, which found the hyperluminous quasar HE 0515-4414 to have a wind luminosity of $\sim0.01\%$ of the bolometric luminosity of the quasar. The $\chi^2$ of our best fitting models is similar to zero signal. Further fitting of halo and feedback models is therefore not merited by the data available in this survey. We do not see evidence of a strongly peaked central decrement that would be indicative of a gNFW profile or feedback wind bubble. We note that the simulations of \citet{Chakraborty2023} show that strong jet-mode feedback is effective at suppressing the SZE from both the halo and thermal winds, thus jet feedback may be occurring in these objects. Only one of the quasars (Q1228+3128, which was not included in our stacking analysis) is currently radio loud, however there could be intermittent jet activity \citep[e.g.][]{2020ApJ...905...74N} and it has been show that non radio-loud quasars can still have radio mode feedback that effect the QSO environment on scales of $\sim$10-100 kpc (\citealt{VillarMartin2021}).

\section{Acknowledgments}
This paper makes use of the ALMA data: 2019.1.01251.S. ALMA is a partnership of ESO (representing its member states), NSF (USA) and NINS (Japan), together with NRC (Canada), MOST and ASIAA (Tai-wan), and KASI (Republic of Korea), in cooperation with the Republic of Chile. The Joint ALMA Observatory is operated by ESO, AUI/NRAO and NAOJ. The National Radio Astronomy Observatory is a facility of the National Science Foundation operated under cooperative agreement by Associated Universities, Inc. We thank the North American ALMA Science Center faculty and staff for their support. We thank the Arizona State University SZ science group, including Seth Cohen, Phil Mauskopf, Sean Bryan, Jenna Moore, Emily Lunde, Jeremey Meinke, Skylar Grayson and Evan Scannapieco, for their advice and contributions.

\bibliography{refs}{}
\end{document}